\documentclass[prl,aps,twocolumn,amsmath,amssymb,floatfix,showpacs,
superscriptaddress]{revtex4}

\usepackage[dvips]{graphics}
\usepackage{color}
\definecolor{dred}{rgb}{0.6,0,0}

\begin{document}

\title{\textcolor{dred}{Spontaneous Photoemission From Metamaterial
Junction: A Conjecture}}

\author{Subir Ghosh}

\email{subir$_$ghosh2@rediffmail.com}

\affiliation{Physics and Applied Mathematics Unit, Indian Statistical
Institute, 203 Barrackpore Trunk Road, Kolkata-700 108, India}

\author{Santanu K. Maiti}

\email{santanu.maiti@isical.ac.in}

\affiliation{Physics and Applied Mathematics Unit, Indian Statistical
Institute, 203 Barrackpore Trunk Road, Kolkata-700 108, India}

\begin{abstract}

The possibility of spontaneous photon pair emission from a normal
material - metamaterial junction is investigated in a quantum field
theory setting. We consider a pair of photons arising from vacuum
fluctuations of the electromagnetic field close to the junction where
one photon each comes from the normal and metamaterial sectors. Mixing 
between the positive and negative norm photon modes can give rise to 
spontaneous photoemission, the rate of which is calculated.

\end{abstract}


\maketitle

In the present work we conjecture the possibility of spontaneous photon
pair creation in normal material - metamaterial junction. Our computational
framework is based on perturbative quantum field theory.
In recent years, metamaterials~\cite{vas}, a novel form of artificially
designed material with electric permittivity $\epsilon$ and magnetic
permeability $\mu$ both being negative, have created a lot of interest
among experimentalists as well as theorists. It has many peculiar
(with respect to conventional material) properties such as negative
refractive index, negative phase velocity, among others that give rise
to a plethora of observational consequences, such as reverse Doppler
effect, reverse Cerenkov effect, perfect lensing, electromagnetic cloaking
and many others~\cite{vas,ex}. 

It is quite natural that normal material - metamaterial interface will
have many unique features, such as negative refraction, superlensing,
anti-parallel directions of phase velocity and Poynting vector,
unconventional surface waves, among others~\cite{jnc1,jnc2,jnc3,jnc4,jnc5}.
However, most of the developments on the theoretical side have been in
the classical electromagnetic wave (EMW) theory framework where
consequences of negative $\epsilon$ and $\mu$ are derived from Maxwell's
equations. Essentially these are refinements of the original results of
Veselgo~\cite{vas}. But, a quantum mechanical or field theoretic study
of metamaterials has not been attempted, to the best of our knowledge. 
In the present paper we show that the counterintuitive features of 
metamaterial, in particular negative phase velocity, can lead to a novel 
phenomenon - {\em photon pair production in conventional positive
(refractive) index material (PIM) - negative (refractive) index material
metamaterial (NIM) junction}. We refer this proposed new effect as
photoemission from metamaterial junction (PMJ).

Many of the observational consequences of NIM appear when dealing with
PIM-NIM junction because of the boundary conditions involved, when
electromagnetic wave crosses the boundary, in a purely classical
setting~\cite{nat}. This feature has proved to be very worthwhile in
Analog Gravity models, after the seminal work of Unruh~\cite{unruh0,rev1}.
In the latter one tries to simulate (theoretically) gravitation like 
features in condensed matter or classical fluid systems and study 
theoretical predictions of gravitation in
experiments that can be performed in the laboratory. In general, signatures
of many of the interesting predictions of gravity, such as Hawking radiation
from Black Holes, gravitational waves, etc., are extremely weak and their
conclusive detection is very difficult. That is why one looks for similar
effects in Analogue Gravity models. In a series of very interesting works,
Smolyaninov and Narimanov~\cite{smol1,smol2} have analyzed the possibility
of simulating many exotic space-time behaviors, governed by gravitation,
such as effective Hawking radiation~\cite{smol2,misn}, metric signature
changing events~\cite{wein}, ``two-time" physics~\cite{bars1,bars2} and
even ``end of time" scenarios~\cite{smol2}, in the framework PIM-NIM 
composites. The analogue gravity concept comes in to play since it is well 
known~\cite{land}
that the dynamical equations governing the EMW in a curved space-time is
similar to the EMW dynamics in PIM with inhomogeneous refractive index.
There is a specific mapping between the metric coefficients on one hand and
permittivity and permeability components on the other. Hence the possibility
of EMW going from PIM to NIM through a junction can lead to metric signature
change in the Analogue Gravity sector where this can amount to time and space
components being interchanged, or even vanishing of conventional time, as one
crosses the (PIM-NIM) interface. In the classical EMW theory, singularities
in the electric field are encountered as one crosses the PIM-NIM
junction~\cite{nat}, which have been observed experimentally~\cite{ex1}. In
the works of~\cite{smol1,smol2} this type of phenomena have been identified
as infinite number of particle (and energy) creation in signature changing
events that have been predicted earlier~\cite{wein} in quantum field theory
computations. However we feel that this identification is too naive and
premature since the field singularities are purely classical in nature and
interpreting this effect as particle (photon) creation, a totally quantum
field theory (QFT) effect, requires further analysis. For this identification
to hold forth rigorously one needs to study the (possibility of) photon
creation in a quantum field theoretic setting. Precisely this has been
attempted in the present paper.

A perturbative QFT scheme for photon pair production where
$\epsilon(\vec{x},t)$ undergoes a space and time varying perturbation
has been provided by Schutzhold {\em et al.}~\cite{sch}. It has been
successfully applied in a specific problem by Belgiorno~\cite{bel} {\em et
al.} where the $\epsilon$ perturbation is produced by a laser pulse
inducing non-linear Kerr effect in the medium. We follow the notation of
Belgiorno {\em et al.}~\cite{bel}. The generic form of amplitude
$A_{(\vec{k},\mu;\vec{k}^{\prime},\mu^{\prime})}$ for a vacuum to a photon
pair transition is,
\begin{equation}
A_{(\vec{k},\mu;\vec{k}^{\prime},\mu^{\prime})}=
\langle\left(\vec{k},\mu;\vec{k}^{\prime},\mu^{\prime}\right)|S| 0\rangle
\label{am}
\end{equation}
where the photon pair is labeled by momenta $\vec{k},\vec{k}^{\prime}$
and polarizations $\mu,\mu^{\prime}$ and the $S$-matrix at first order of
$\epsilon$-perturbation is $S\sim 1-i\int d^4x {\mathcal{H_I}(x)}$. The
interaction Hamiltonian density is defined as
${\cal{H_I}}(x)=\xi \vec D(\vec x,t)^2$, where
\begin{equation}
\xi=\frac{1}{2}\left(\frac{1}{n^2 (\vec x)}-\frac{1}{(n_0)^2}\right).
\label{int}
\end{equation}
$\vec{D}(\vec{x},t)$ is the displacement vector (or equivalently the
canonical momentum~\cite{sch,bel}) and $\epsilon(\vec x)=n^2(\vec x)$ and
$\epsilon_0=n^2_0$ are the variable and constant background refractive
indices, respectively. Indeed in these applications only PIMs are considered.
For a non-zero amplitude to exist, {\em it is essential for the perturbation
to be explicitly time-dependent}~\cite{sch,bel}. This follows simply from
the principle of energy conservation which for a static perturbation will
appear as $\delta(\omega_k+\omega_{k^{\prime}})$ in a generic transition
amplitude $A_{(\vec{k},\mu;\vec{k}^{\prime},\mu^{\prime})}$ and can never
be satisfied for $+$ve $\omega$ for PIM. For this reason a time-dependent
$\epsilon(\vec{x},t)$ perturbation is necessary for a non-vanishing amplitude
since it introduces additional terms in the energy conservation
$\delta$-function that helps to saturate the $\delta$-function (for details
see~\cite{bel}).

But for a NIM such as metamaterial things are radically different. Inside a 
PIM medium $\omega_k=ck/n_0, k=|\vec{k}|$ and the right (left) moving plane 
waves (in one spatial dimension) are represented by $\sim \exp(ikx-i\omega t)$ 
($\sim \exp(-ikx-i\omega t)$). In NIM one can simulate the negative phase 
velocity by replacing $n_0$ by $-n_0$ which converts the above waves to
$\sim \exp(ikx+i\omega t)$ and $\sim \exp(-ikx+i\omega t)$. Notice that the 
NIM states turn out to be complex conjugates of the PIM states and hence 
are negative norm states whereas the PIM states are positive norm states. 
This mode-mixing phenomenon~\cite{unruh} at the PIM-NIM junction can give 
rise to our 
conjectured PMJ effect. It should be stressed that negative phase velocity 
of NIM modes have striking physical consequences such as negative refraction, 
reverse Cerenkov effect, reverse Doppler effect, etc., that are already 
established.

We propose the following scenario: in a vacuum quantum fluctuation {\em a
photon pair is produced in a PIM-NIM junction such that one photon is in
the PIM sector and the other is in the NIM sector}. Then, even for a time
independent perturbation, the $\delta$-function will appear as
$\delta(\omega^{\mbox{\tiny PIM}}_k-\omega^{\mbox{\tiny NIM}}_{k^{\prime}})$
that can be saturated yielding a non-zero vacuum to photon pair transition
amplitude. Still a question remains as to from where will the energy come
for the photon pair production. Note that throughout our analysis we have
ignored dispersion effects and loss indicating that we are restricted to 
a narrow frequency band. Without dispersion the NIM is unstable with 
negative energy density and losslessness leads to violation of
\begin{figure}[ht]
{\centering \resizebox*{5.75cm}{3.25cm}{\includegraphics{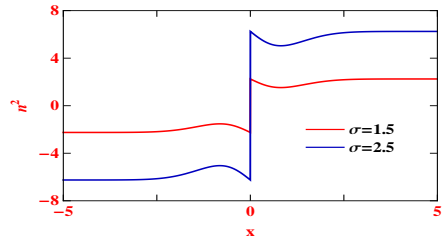}}\par}
\caption{(Color online). $x$-dependence of $n^2(\vec{x})$ for two different
values of $\sigma$.}
\label{fig1}
\end{figure}
Kramers-Kronig relation. Hence energy from outside is needed to stabilize 
the NIM and this energy comes out in the PMJ. A somewhat similar 
idea was suggested in~\cite{smol2} that energy needs to be supplied 
from outside to compensate for the dissipation in NIM, which we have not 
considered. The main point is that, in a dispersionless and lossless model 
(as considered in many previous works), PMJ can occur in a static PIM-NIM 
junction with no time-dependent~\cite{sch} (eg. laser induced~\cite{bel})
$\epsilon$-perturbation in the process. Indeed it is imperative to study 
effects of dispersion and loss on the PMJ. 

We notice a qualitative resemblance between PMJ and the celebrated 
Hawking-Unruh effect of Black Hole evaporation~\cite{haw1,unruh1} or more 
closely with Hawking-Unruh radiation in Analogue Gravity 
scenario~\cite{unruh0}. In the semi-classical explanation of Hawking effect, 
out of the photon pair (or any other particle pair) created close to the 
event horizon, the negative energy particle is trapped inside 
the Black Hole (thereby reducing the Black Hole's mass) whereas its positive 
energy partner escapes. The latter constitutes the Hawking radiation. Energy 
conservation is taken care of since the Black Hole mass reduces. In the 
Analogue Gravity scenario this effect is captured by mode-mixing between 
positive and negative norm states~\cite{unruh}. The  major differences 
between Hawking effect and PMJ are that (i) in PMJ, both photons of the 
correlated pair should be observable; (ii) the energy needed in PMJ has 
to be provided from outside to preserve the NIM and (iii) only photon pairs 
are involved in PMJ whereas Hawking-Unruh radiation, in principle, can 
constitute of all types of elementary particles. In an Analogue Gravity 
scenario, Belgiorno 
{\em et al.}~\cite{bel1} have claimed to provide evidence of analogue 
Hawking radiation in a controllable moving refractive index perturbation 
using ultrashort Laser pulse filamentation~\cite{fac}. However the above 
claim~\cite{bel1} has been debated in~\cite{bel3}.

We now attempt to provide a quantitative estimate of PMJ. In general it is
tricky to consider a perturbation small if it incorporates a change of sign
in the physical parameter in question, as is true for the present case. We
have considered the composite material such that the negative $x$ (positive
$x$) consists of NIM (PIM) of constant background values $-n^2_0$ ($+n^2_0$)
and the perturbation smooths out the crossover at the junction. Hence the
overall refractive index is given by,
\begin{equation}
n^2 (\vec{x})=n_0^2\,\mbox{sign}(x)-2 n_0\eta \tanh(x/l)e^{-R^2/(2\sigma^2)},
\label{ep}
\end{equation}
where, $\mbox{sign}(x)=+(-)1$ for $x>(<)0$, $\eta$ is a small parameter and
$l,\sigma$ describe the crossover at the junction and $R^2=x^2+y^2+z^2$.
\begin{figure}[ht]
{\centering \resizebox*{7cm}{6cm}{\includegraphics{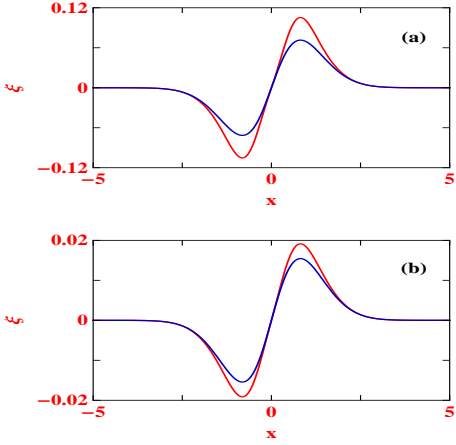}}\par}
\caption{(Color online). The variation of $\xi$ as a function of $x$, where
(a) $\sigma=1.5$ and (b) $\sigma=2.5$. The red and blue lines correspond to
the exact and approximated expressions given in Eq.~\ref{xi}, respectively.}
\label{fig2}
\end{figure}
At large distance from the junction at $x=0$ the left and right side
materials are NIM and PIM, respectively. This is depicted in Fig.~\ref{fig1}
for two values of $\sigma$, where the red and blue lines correspond to
$\sigma=1.5$ and $2.5$, respectively. It should be noted that these values
of $\sigma$ are taken only to demonstrate the behavior of $n^2(\vec{x})$
pictorially whereas in our numerical computation later we have taken a
realistic value of $\sigma=1.5\,\mu \mbox{m}$. Now to $O(\eta)$ we can write,
\begin{eqnarray}
\xi & = & \frac{1}{2}\left[\frac{1}{\mbox{sign}(x)\,n_0^2 - 2\,n_0\,\eta
\tanh(x/l)\,e^{-R^2/(2\sigma^2)}} \right. \nonumber \\
& & \left. - \frac{1}{\mbox{sign}(x)\,n_0^2}\right]
\nonumber \\
 & \approx & \frac{\eta\,\tanh(x/l)e^{-R^2/(2\sigma^2)}}
{(\mbox{sign}(x))^2n_0^3}\nonumber \\
 & = & \frac{\eta\,\tanh(x/l)e^{-R^2/(2\sigma^2)}}{n_0^3}.
\label{xi}
\end{eqnarray}
In the last step we have used $(\mbox{sign}(x))^2=1$ from Generalized
Function perspective (see for example~\cite{gen}). The justification of this
prescription is illustrated in Fig.~\ref{fig2} where we plot the exact function
(red curve) given in the equality Eq.~\ref{xi}, and its approximate version
(blue curve) that we actually use given in the third equality in Eq.~\ref{xi}.
The results are presented for two different values of $\sigma$, for the sake
of simplicity.

The Fourier transform turns out to be~\cite{tilde},
\begin{eqnarray}
\tilde{\xi} & = & \frac{\eta}{n_0^3} \int dt\,dx\,dy\,dz
\tanh(x/l)e^{-R^2/(2\sigma^2)} \nonumber \\
 & = & \frac{8\pi^2 \eta \sigma^4 k_x \delta(\omega_k)}{n_0^3}
e^{-\frac{\sigma^2}{2}\vec{k}^2}.
\label{til}
\end{eqnarray}
The photoemission number becomes,
\begin{eqnarray}
N_{k,\mu} & = & \frac{V}{(2\pi)^3}\int d^3k^{\prime}
\frac{\omega_k\omega_{k^{\prime}}}{V^2} |\tilde{\xi}(k+k^{\prime})|^2
\left[1-(\vec{e}_k.\vec{e}_{k^{\prime}})^2\right] \nonumber \\
 & = & \frac{16\pi^2L\eta^2\sigma^6k}{Vn_0^6}\int d^3k^{\prime}\,
\delta(k-k^{\prime})\sigma^2(k_x+k^{\prime}_x)^2 \nonumber \\
 & & \times e^{-\sigma^2 |k+k^{\prime}|^2}
\left[1-(\vec{e}_k.\vec{e}_{k^{\prime}})^2\right].
\label{n}
\end{eqnarray}
The energy contributions in the $\delta$-function appear with opposite
signature which, as explained before, is the highlight of our conjecture.
\begin{figure}[ht]
{\centering \resizebox*{6cm}{6cm}{\includegraphics{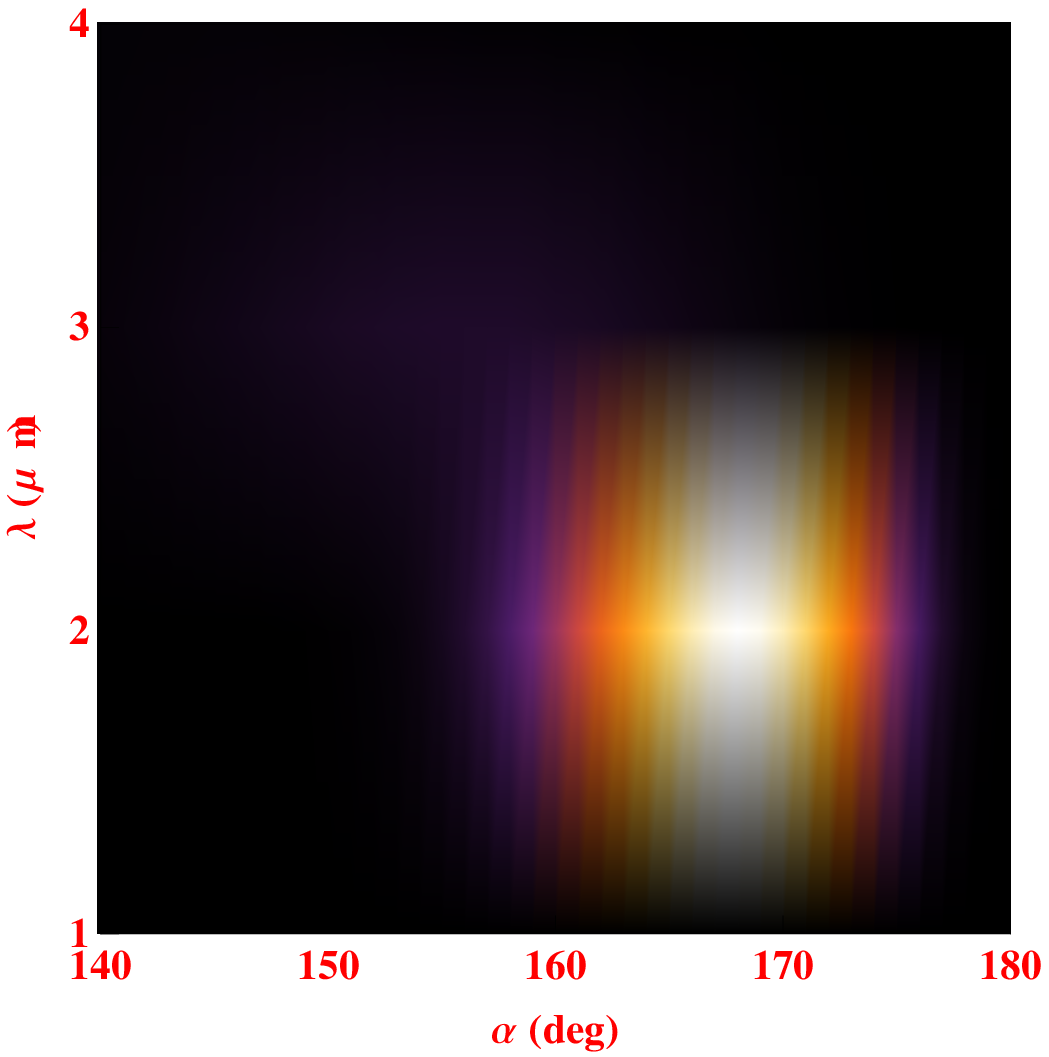}}
\raise 0.6cm\hbox{\kern 0.5cm\resizebox*{0.7cm}{5.35cm}
{\includegraphics{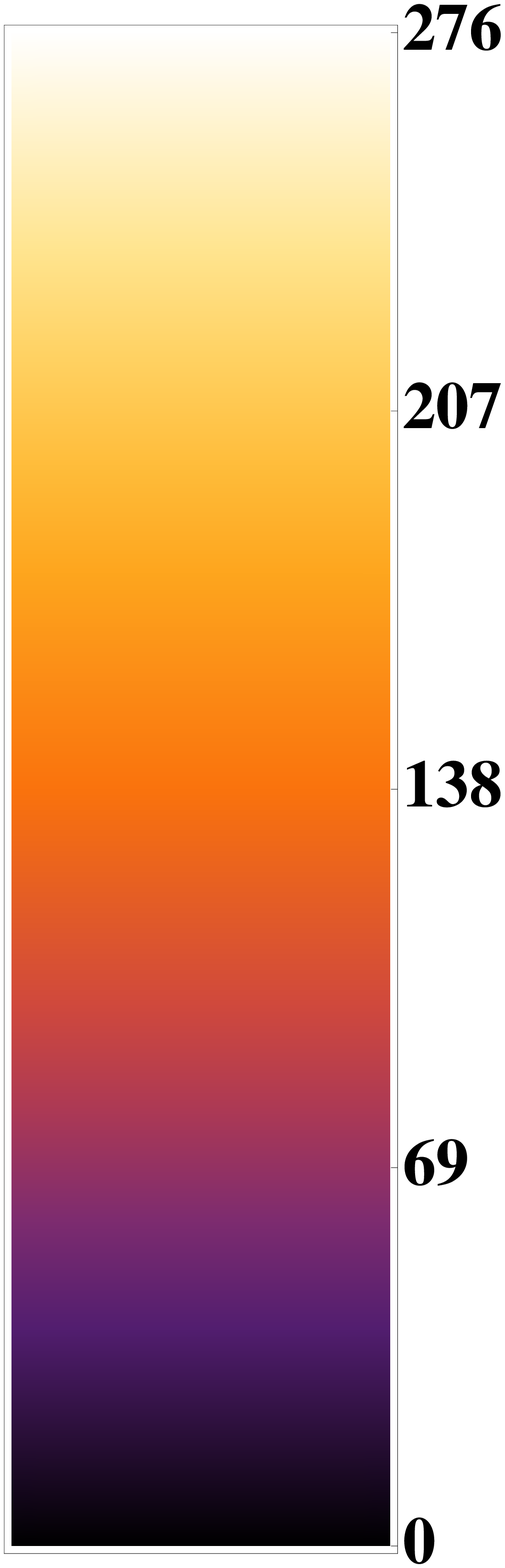}}}\par}
\caption{(Color online). The rate of photoemission $dN/d\Omega$ computed
from Eq.~\ref{ddn} for different values of $\alpha$ and $\lambda$ when we
set $\sigma=1\,\mu \mbox{m}$ and $n_0=2.5$.}
\label{fig3}
\end{figure}
Rest of the calculation is straightforward. Defining
$(\vec{k}^{\prime})^2=(k_x^{\prime})^2+r^2$, the result becomes
\begin{eqnarray}
N_{k,\mu} & = & \frac{16 \pi^2 L \eta^2 \sigma^8 k}{Vn_0^6} \int
\limits_0^{\infty} dr \int \limits_0^{2\pi} d\theta \frac{r (f^2+r^2)}{f}
\nonumber \\
 & & \times (k_x+f)^2 \frac{\left(k_xf+k_{\bot} r \cos\theta\right)^2}
{k^2(f^2+r^2)} \nonumber \\
 & & \times e^{-\sigma^2\left[(k_x+f)^2 + k_{\bot}^2 + r^2 +
2k_{\bot}r\cos\theta\right]}
\label{dn}
\end{eqnarray}
where $f(r)={\sqrt{\vec{k}^2-r^2}}$. The differential cross section is
given by,
\begin{eqnarray}
\frac{dN}{d\Omega} & = & \frac{64 \pi^3 L \eta^2 \sigma^8}{n_0^6} k dk
\int \limits_0^{\infty} dr \int \limits_0^{2\pi} d\theta
\frac{r}{f}(k_x+f)^2 \nonumber \\
 & \times & (k_xf+k_{\bot} r \cos\theta)^2
e^{-\sigma^2\left[(k_x+f)^2 + k_{\bot}^2 + r^2 + 2k_{\bot}r\cos\theta\right]}.
\nonumber \\
\label{ddn}
\end{eqnarray}
This constitutes our main result. Figure~\ref{fig3} describes a numerical
evaluation of our result (Eq.~\ref{ddn}) for $n_0=2.5$, $\eta=0.01$, and
$\sigma =1\,\mu \mbox{m}$. The photoemission number density is plotted
as functions wavelength $\lambda$ and emission angle $\alpha$. We notice
that the process is fairly localized in $\alpha$ and $\lambda$ with a
maximum emission at around $\lambda =2\,\mu \mbox{m}$. The maximum count
rate is $\sim200$ but indeed the number should not be taken too seriously
at our present level of analysis.

We conclude that the possibility of a completely new form of spontaneous 
photoemission is suggested from a normal material - metamaterial interface. 
The junction acts as a ``horizon" that separates two types of excitations 
with opposite phase velocities. This makes spontaneous emission of a photon 
pair feasible due to mixing between positive and negative norm modes, 
provided one of the photons comes from the normal material side of the 
junction and the other from the metamaterial side. The proposed phenomenon 
is reminiscent of Hawking-Unruh radiation in Analog Gravity framework. 
However, to truly establish this new form of spontaneous photoemission it 
is imperative that one takes dispersive effects into account. Studies are 
being pursued in this direction. 

It is indeed a great pleasure to thank F. Belgiorno and D. Faccio for
helpful correspondences. We are very grateful to W. G. Unruh and 
I. I. Smolyaninov for suggestions.


\begin{thebibliography}{99}
\bibitem{tilde} An exponential damping term in the Fourier integral of
$x$-sector is dropped so that the Fourier transform is computed analytically.
Note that part of the contribution of the exponential term in $x$ appears in
$\exp[-(\sigma^2k_x^2)/2]$ in Eq.~\ref{til}. The above simplification can
affect our numerical estimates of photon emission number.
\end{thebibliography}
\end{document}